\documentclass{llncs}
\usepackage{graphicx}
\usepackage{color}
\usepackage{align-objs}
\usepackage{mathrsfs}
\usepackage{amsmath}
\usepackage{amsfonts}
\usepackage{amssymb}
\usepackage{todonotes}
\usepackage{eucal}
\usepackage{eufrak}

\newcommand{\spth}{\rightsquigarrow}
\newcommand{\ZI}{Z^I}
\newcommand{\ZIprime}{Z^{I'}}

\newcommand{\ZB}[1]{Z^{B,#1}}
\newcommand{\YB}[1]{Y^{B,#1}}

\newcommand{\Red}[1]{\textcolor{red}{#1}}
\newcommand{\ignore}[1]{}
\DeclareMathOperator{\lcd}{lcd}

\spnewtheorem*{proofsketch}{Proof (sketch)}{\itshape}{\rmfamily}

\begin{document}

\markboth{}
{Graph-Distance within RNA Secondary Structure Ensembles}

\title{DISTRIBUTION OF GRAPH-DISTANCES IN
BOLTZMANN ENSEMBLES OF RNA SECONDARY STRUCTURES}

\author{Rolf Backofen\inst{1-2}\and Markus Fricke\inst{3}\and Manja
  Marz\inst{3} \and \\ Jing
  Qin\inst{4} \and  Peter F. Stadler\inst{4-8}}

\institute{Department of Computer Science,Chair for Bioinformatics,
  University of Freiburg, Georges-Koehler-Allee 106, D-79110 Freiburg,
\and
Center for Biological Signaling Studies (BIOSS), Albert-Ludwigs-Universit\"{a}t,
Freiburg, Germany
\and
Bioinformatics/High Throughput Analysis
Faculty of Mathematics und Computer Science
Friedrich-Schiller-University Jena
Leutragraben 1, 07743 Jena
\and
  Max Planck Institute for Mathematics in the Sciences,
  Inselstra{\ss}e 22, 04103 Leipzig, Germany
\and
Bioinformatics Group, Department of Computer Science,
  and Interdisciplinary Center for Bioinformatics,
  University of Leipzig,
  H{\"a}rtelstrasse 16-18, 04107 Leipzig, Germany
\and
  Fraunhofer Institut for Cell Therapy and
  Immunology, Perlickstra{\ss}e 1,04103 Leipzig,
  Germany
\and
  Institute for Theoretical Chemistry, University of Vienna,
  W{\"a}hringerstrasse 17, A-1090 Vienna, Austria
\and
  Santa Fe Institute, 1399 Hyde Park Rd., Santa Fe, NM87501, USA.
}

\maketitle

\begin{abstract}
  Large RNA molecules often carry multiple functional domains whose spatial
  arrangement is an important determinant of their function. Pre-mRNA
  splicing, furthermore, relies on the spatial proximity of the splice
  junctions that can be separated by very long introns. Similar effects
  appear in the processing of RNA virus genomes. Albeit a crude measure,
  the distribution of spatial distances in thermodynamic equilibrium therefore
  provides useful information on the overall shape of the molecule can
  provide insights into the interplay of its functional domains. Spatial
  distance can be approximated by the graph-distance in RNA
  secondary structure. We show here that the equilibrium
  distribution of graph-distances between arbitrary nucleotides can be computed
  in polynomial time by means of dynamic programming. A naive
  implementation would yield recursions with a very high time complexity of
  $O(n^{11})$. Although we were able to reduce this to $O(n^6)$ for many
  practical applications a further reduction seems difficult. We conclude,
  therefore, that sampling approaches, which are much easier to implement,
  are also theoretically favorable for most real-life applications, in
  particular since these primarily concern long-range interactions in very
  large RNA molecules.
\end{abstract}

\section{Introduction}

The distances distribution within an RNA molecule is of interest in various
contexts.  Most directly, the question arises whether panhandle-like
structures (in which 3' and 5' ends of long RNA molecules are placed in
close proximity) are the rule or an exception. Panhandles have been
reported in particular for many RNA virus genomes. Several studies
\cite{Yoffe:11,Fang:11,Clote:12,Han:12} agree based on different models
that the two ends of single-stranded RNA molecules are typically not far
apart. On a more technical level, the problem to compute the partition
function over RNA secondary structures with given end-to-end distance $d$,
usually measured as the number of external bases (plus possibly the number
of structural domains) arises for instance when predicting nucleic acid
secondary structure in the presence of single-stranded binding proteins
\cite{Forties:10} or in models of RNA subjected to pulling forces (e.g.\ in
atom force microscopy or export through a small pore)
\cite{Gerland:01,Mueller:02,Gerland:04}. It also plays a role for the
effect of loop energy parameters \cite{Einert:08}.

In contrast to the end-to-end distance, the graph-distance between two
\textit{arbitrarily} prescribed nucleotides in a larger RNA structure does
not seem to have been studied in any detail. However, this is of particular
interest in the analysis of single-molecule fluorescence resonance energy
transfer (smFRET) experiments \cite{Roy:2008}. This technique allows to
monitor the distance between two dye-labeled nucleotides and can reveal
details of the kinetics of RNA folding in real time. It measures the
non-radiative energy transfer between the dye-labeled donor and acceptor
positions. The efficiency of this energy transfer, $E_{fret}$, strongly
depends on the spatial distance $R$ according to
$E_{fret}=(1+(R/R_0)^6)^{-1}$. The F{\"o}rster radius $R_0$ sets the length
scale, e.g.\ $R_0\approx 54$ \AA{} for the Cy3-Cy5 dye pair. A major
obstacle is that, at present, there is no general and efficient way to link
smFRET measurements to interpretations in terms of explicit molecular
structures.  To solve this problem, a natural first step to compute the
distribution of spatial distances for an equilibrium ensemble of 3D
structures. Since this is not feasible in practice despite major progress
in the field of RNA 3D structure prediction \cite{Das:07}, we can only
resort to considering the graph-distances on the ensemble of RNA secondary
structures instead.  Although a crude approximation of reality, our initial
results indicate that the graph distance can be related to the smFRET data
such as those reported by \cite{Kobitski:07}. From a computer science point
of view, furthermore, we show here that the distance distribution can be
computed exactly using a dynamic programming approach.

\section{Theory}
\subsection{RNA Secondary Structures}

An RNA secondary structure is a vertex labeled outerplanar graph $G(V,\xi,
E)$, where $V=\{1,2,\dots,n\}$ is a finite \emph{ordered} set (of
nucleotide positions) and $\xi:\{1,2,\dots,n\}\to
\{\mathsf{A},\mathsf{U},\mathsf{G},\mathsf{C}\}, i\mapsto \xi_i$ assigns to
each vertex at position $i$ (along the RNA sequence from 5' to 3') the
corresponding nucleotide $\xi_i$. We write $\xi=\xi_1\ldots\xi_n$
for the \emph{sequence} underlying secondary structure and use 
$\xi[i\ldots j]=\xi_i\ldots\xi_j$ to denote the \emph{subsequence} 
from $i$ to $j$. The edge set $E$ is subdivided into backbone edges of the
form $\{i,i+1\}$ for $1\le i <n$ and a set $B$ of base pairs satisfying the
following conditions:
\begin{enumerate}
\item If $\{i,j\}\in B$ then $\xi_i\xi_k \in \{ \mathsf{GC}, \mathsf{CG},
  \mathsf{AU}, \mathsf{UA}, \mathsf{GU}, \mathsf{UG} \}$.
\item If $\{i,j\}\in B$ then $|j-i|>3$. 
\item If $\{i,j\}, \{i,k\} \in B$ then $j=k$
\item If $\{i,j\}, \{k,l\} \in B$ and $i<k<j$ then $i<l<j$. 
\end{enumerate}
The first condition allows base pairs only for Watson-Crick and GU base
pairs. The second condition implements the minimal steric requirement for
an RNA to bend back on itself. The third condition enforces that $B$ forms a
matching in the secondary structure. The last condition (nesting condition) 
forbids crossing base pairs, i.e.\ pseudoknots. 

The nesting condition results in a natural partial order in the set of
base pairs $B$ defined as $\{i,j\}\prec\{k,l\}$ if $k<i<j<l$. In
particular, given an arbitrary vertex $k$, the set $B_{k}=\{ \{i,j\}\in B |
i\le k\le j\}$ of base pairs enclosing $k$ is totally ordered. Note that
$k$ is explicitly allowed to be incident to its enclosing base pairs. A
vertex $k$ is \emph{external} if $B_{k}=\emptyset$. A base pair $\{k,l\}$
is \emph{external} if $B_{k}=B_l=\{ \{k,l\}\}$.

Consider a fixed secondary structure $G$, for a given base pair $\{i,j\}\in
B$, we say a vertex $k$ is \emph{accessible} from $\{i,j\}$ if $i<k<j$ and
there is no other pair $\{i',j'\}\in B$ such that $i<i'<k<j'<j$. The unique
subgraph $\mathcal{L}_{i,j}$ induced by $i$, $j$, and all the vertices
accessible from $\{i,j\}$ is known as the \emph{loop} of $\{i,j\}$.  The
\emph{type} of a loop $\mathcal{L}_{i,j}$ \ignore{as one of six categories
  [(1) hairpin loop; (2) stacked pair; (3) bulge loop; (4) interior loop;
  (5) multiloop or (6) exterior loop]} is unique determined depending on
whether $\{i,j\}$ is external or not, and the numbers of unpaired vertices
and base pairs. For details, see
\cite{Schuster:Fontana:Stadler:From_seque_shape:1994}. Each secondary
structure $G$ has a unique set of loops $\{\mathcal{L}_{i,j}\vert
\{i,j\}\in B\}$, which is called the \emph{loop decomposition} of $G$. The
free energy $f(G)$ of a given secondary structure, according to the
standard energy model \cite{Mathews:04a}, is defined as the sum of the
energies of all loops in its unique loop decomposition.

The relative location of two vertices $v$ and $w$ in $G$ is determined by
the base pairs $B_{v}$ and $B_{w}$ that enclose them. If $B_{v}\cap
B_{w}\ne \emptyset$, there is a unique $\prec$-minimal base pair
$\{i_{v,w},j_{v,w}\}$ that encloses both vertices and thus a uniquely
defined loop $\mathcal{L}_{\{i_{v,w},j_{v,w}\}}$ in the loop associated
with $v$ and $w$.  If $B_{v}\setminus B_{w}=\emptyset$ or $B_{w}\setminus
B_{v}=\emptyset$ then $v$ or $w$ is unpaired and part of
$\mathcal{L}_{\{i_{v,w},j_{v,w}\}}$.  Otherwise, i.e.~$B_{v}\cap
B_{w}=\emptyset$, there are uniquely defined $\prec$-maximal base pairs
$\{k_v,l_v\}\in B_{v}\setminus B_{w}$ and $\{k_w,l_w\}\in B_{w}\setminus
B_{v}$ that enclose $v$ and $w$, respectively. This simple partition holds
the key to computing distance distinguished partition functions below.

It will be convenient in the following to introduce edge weights
$\omega_{i,j}= a$ if $j=i+1$, i.e., for backbone edges,
and $\omega_{i,j}=b$ for $\{i,j\}\in B$. Given a path $p$, we define
the weight of the path $d(p)$ as the sum of the weights of edges in the path. 
The (weighted) \emph{graph-distance} 
$d^G_{v,w}$ in $G$ is defined as the
weight of the path $p$ connecting $v$ and $w$ with $d(p)$ being minimal. 
For the weights, we require the following condition:
\begin{itemize}
  \item[(W)] If $i$ and $j$ are connected by an edge, then $\{i,j\}\in E$
    is the unique shortest path between $i$ and $j$. 
\end{itemize}
This condition ensures that single edges cannot be replaced by detours of
shorter weight. Condition (W) and property (ii) of the secondary structure
graphs implies $b<3a$ because the closing base pair must be shorter than a
hairpin loop. Furthermore, considering a stacked pair we need $b<b+2a$,
i.e.\ $a>0$. We allow the degenerate case $b=0$ that
neglects the traversals of base pairs.

\subsection{Boltzmann Distribution of Graph-Distances}

For a fixed structure $G$, $d^G_{v,w}$ is easy to compute. Here, we are
interested in the distribution $Pr[d^G_{v,w}|\xi]$ and its expected value
$d_{v,w}=E[d^G_{v,w}|\xi]$ over the ensemble of all possible structures $G$
for a given sequence $\xi$. Both quantities can be calculated from the
Boltzmann distribution $Pr[G|\xi] = e^{-f(G)/RT}/Q$ where $Q=\sum_G
e^{-f(G)/RT}$ denotes the partition function of the ensemble of
structures. As first shown in \cite{McCaskill:1990}, $Q$ and related
quantities can be computed in cubic time.  A crucial
quantity for our task is the restricted partition function
\begin{displaymath}
  Z^{v,w}[d] =\sum_{G\text{ with }d^G_{v,w}=d} e^{-f(G)/RT}
\end{displaymath}
for a given pair $v,w$ of positions in a given RNA sequence $\xi$. A simple
but tedious computation (Appendix A \footnote{The Appendix A-D
  of our paper are available from
  \url{http://www.rna.uni-jena.de/supplements/RNAgraphdist/supplement.pdf}.}) verifies that the
$Pr[d^G_{v,w}=d|\xi]= Z^{v,w}[d]/Q$ and $d_{v,w}=E[d^G_{v,w}|\xi]= \sum_d
(Z^{v,w}[d]/Q)d$. Hence it suffices to compute $Z^{v,w}[d]$ for
$d=1,\dots,n$. In sections \ref{S:Part1}-\ref{S:Part3} we show that this
can be achieved by a variant of McCaskill's approach \cite{McCaskill:1990}.

For the ease of presentation we describe in the following only the
recursion for the simplified energy model for the ``circular maximum
matching'' matching, in which energy contributions are associated with
individual base pairs rather than loops. Our approach easily extends
to the full model by using separating the partition functions into distinct 
cases for the loop types. We use the letter $Z$ to denote partition 
functions with distance constraints, while $Q$ is used for quantities 
that appear in McCaskill's algorithm and are considered as  pre-computed
here. 

Before we continue with the calculation of the partition function, let's
first look into problem formulation in more detail. For the FRET
application, it is well-known that the rate which with FRET occurs is
correlated with the distance.  Therefore, only a limited range of distance
changes (e.g.~ $20\AA-100 \AA$ for Cy3-Cy5) can be reported by the FRET
experiments.  Thus the more useful formulation of our problem is not to use
the full expected quantity for all positions. Instead, we are interested in
the average for all distances within some threshold $\theta_d$.  
As the
space and time complexity will depend on the number of distances we
consider, we will parametrise our complexity by the number of nucleotides
$n$ and the number of overall distances considered $D=\theta_d+1$, as well.

\subsection{Recursions of $Z^{v,w}[d]$: $v$ and $w$ Are External}
\label{S:Part1}

An important special case assumes that both $v$ and $w$ are external. This
is case e.g.\ when $v$ and $w$ are bound by proteins. In particular, the
problem of computing end-to-end distances, i.e., $v=1$ and $w=n$, is of
this type.  Assuming (W), the shortest path between two external vertices
$v,w$ consists of the external vertices and their backbone connections
together with the external base pairs. We call this path the \emph{inside
  path} of $i,j$ since it does not involve any vertices ``outside'' the
subsequence $\xi[i..j]$.

For efficiently calculating the internal distance between any two
vertices $v,w$, we denote by $\ZI_{i,j}[d]$ 
the partition function over all secondary structures on $\xi[i..j]$ with
end-to-end distance exactly $d$. Furthermore, let $Q^B_{i,j}$ denote the 
partition function over all secondary structures  on $\xi[i..j]$ that 
are enclosed by the base pair $\{i,j\}$. We will later also
need the partition function $Q_{i,j}$ over the sub-sequence $\xi[i..j]$,
regardless of whether $\{i,j\}$ is paired or not.

Now note that any structure on $\xi[i..j]$ starts either with an unpaired
base or with a base pair connecting $i$ to some position $k$ satisfying
$i<k\le j$. In the first case, we have
$d^{G}_{i,j}=d^{G}_{i,i+1}+d^{G}_{i+1,j}$ where $d^{G}_{i,i+1}=a$.  In the
second case, there exists
$d^{G}_{i,j}=d^{G}_{i,k}+d^{G}_{k,k+1}+d^{G}_{k+1,j}$ with $d^{G}_{i,k}=b$
and $d^{G}_{k,k+1}=a$.  Thus, $\ZI_{i,j}[d]$ can be split as follows,
 \begin{displaymath}
   \includegraphics[width=0.7\textwidth]{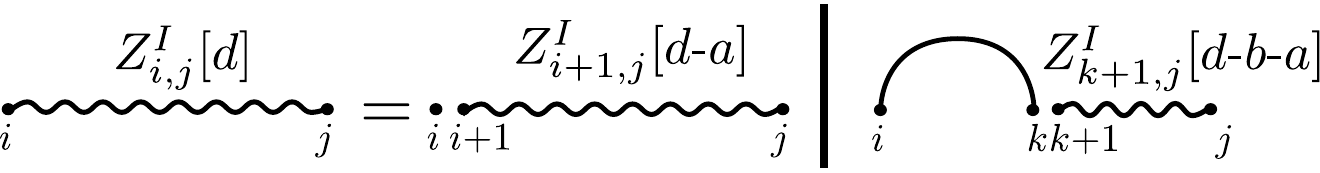}
 \end{displaymath}
This gives the recursion 
\begin{equation}
\label{eq:ZI}
  \ZI_{i,j}[d] = \ZI_{i+1,j}[d-a]  + 
  \sum_{i<k\le j} Q^B_{i,k} \ZI_{k+1,j}[d-b-a]
\end{equation} 
with the initialization $\ZI_{ii}[0]=1$ and $\ZI_{ii}[d]=0$ for $d>0$.  For
consecutive vertices we have $\ZI_{i,i+1}[a]=1$ and $\ZI_{i,i+1}[d]=0$ for
$d\ne a$.  These recursions have been derived in several different
contexts, e.g.\ force induced RNA denaturations \cite{Gerland:01}, the
investigate of loop entropy dependence \cite{Einert:08}, the analysis of 
FRET signals in the presence of single-stranded binding proteins
\cite{Forties:10}, as well as in mathematical studies of RNA panhandle-like
structures \cite{Clote:12,Han:12}. 

In the following it will be convenient to define also a special terms for
the empty structure. Setting $\ZI_{i,i-1}[-a]=1$ and $\ZI_{i,i-1}[d]=0$ for
$d \ne -a$ allows us to formally write an individual backbone edge as two
edges flanking the empty structure and hence to avoid the explicit
treatment of special cases. This definition of $\ZI$ also includes the case
that $i$ and $j$ are base paired in the recursion (\ref{eq:ZI}). This is
covered by the case $k=j$, where we evaluate $\ZI_{j+1,j}[d-b-a]$. Since
$d=b$ is the only admissible value here, this refers to $\ZI_{j+1,j}[-a]$,
which has the correct value of $1$ due to our definition. Later on, we will
also need $\ZI$ under the additional condition that the path starts and end
with a backbone edge. We therefore introduce $\ZIprime$ defined as
\begin{center}
  \includegraphics[width=0.7\textwidth]{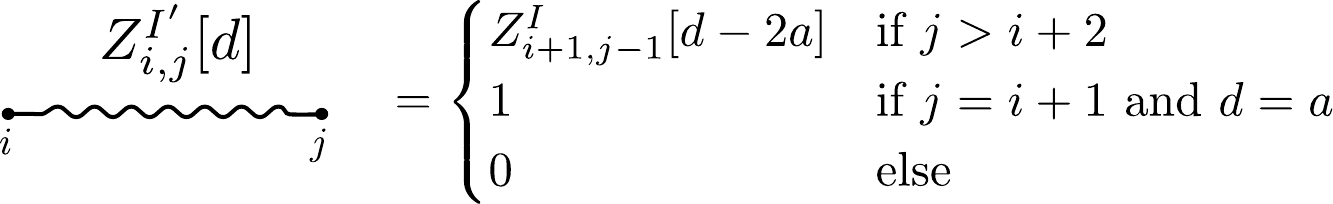}
\end{center}
By our initialization of $\ZI$, we can simply define $\ZIprime$ by
\begin{equation}
\label{eq:ZIprime}
  \ZIprime_{i,j}[d] =  \ZI_{i+1,j-1}[d-2a]
\end{equation}
Note that if $\ZIprime_{i,j}[d]$ is called with $j=i+1$, then we call
$\ZI_{i+1,i}[d-2a]$. The only admissible value again is the correct
value $d=a$.

This recursion requires $O(Dn^3)$ time and space. It is possible to reduce
the complexity in this special case by a linear factor. The trick is to use
conditional probabilities for arcs starting at $i$ or the conditional
probability for $i$ to be single-stranded, which can be determined from the
partition function for RNA folding \cite{Clote:12}, see Appendix B.

\subsection{Recursions of $Z^{v,w}[d]$: The General Case}

\begin{figure}[t]
\begin{tabular}{lr}
\begin{minipage}{0.6\textwidth}
\includegraphics[width=\textwidth]{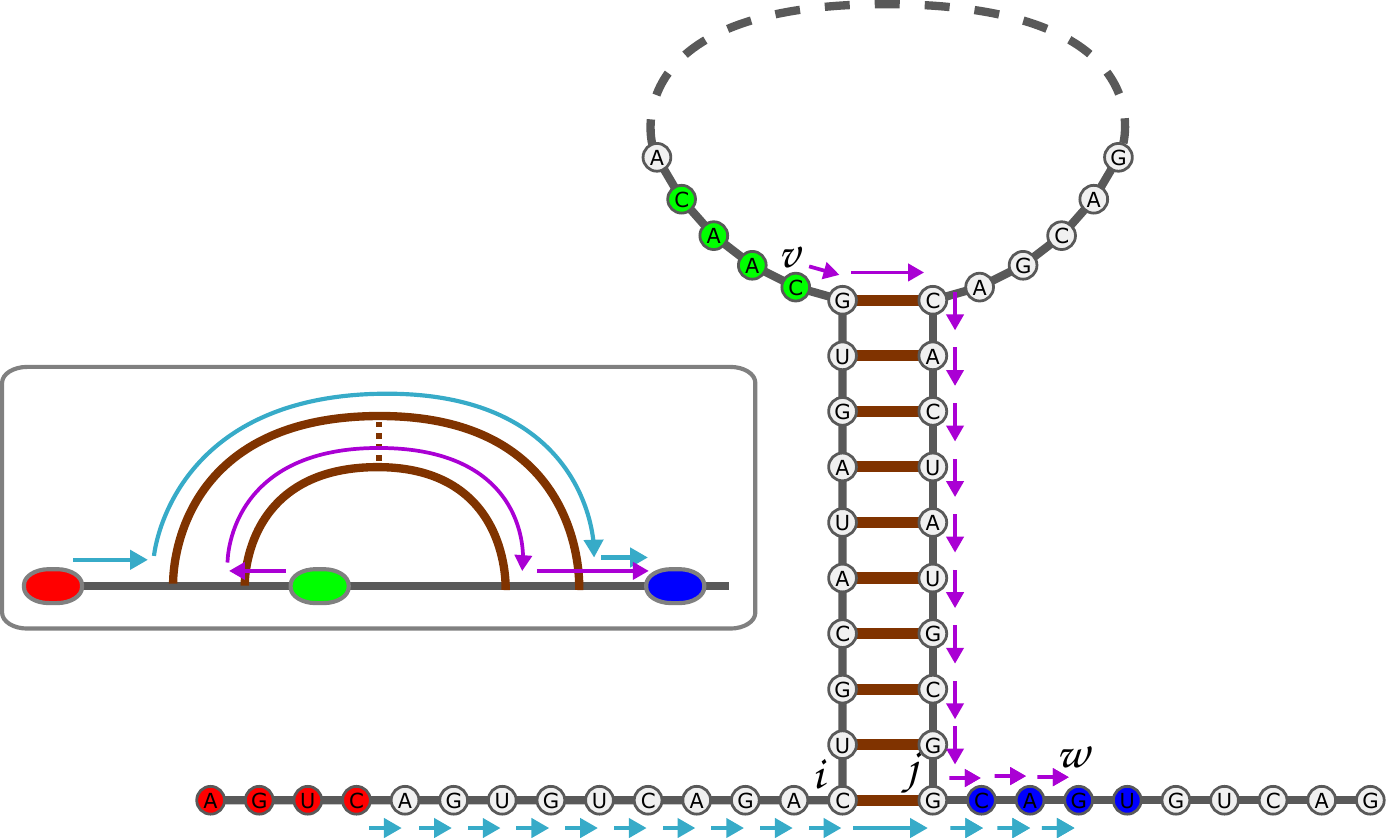}
\end{minipage}
&
\begin{minipage}{0.38\textwidth}
  \caption{Inside and outside paths. 
    The shortest path (violet arrows) from $v$ (green) to $w$ (blue)  
    is not an inside path: \emph{inside} emphasizes that, in contrast 
    to the shortest path (cyan arrows) between the red region and $w$, 
    it is not contained in the interval determined by its end points.}
\label{fig:RNA-two-graphs}
\end{minipage}
\end{tabular}
\end{figure}

The minimal distance between two positions that are covered by an arc can
be realized by \emph{inside paths} and \emph{outside paths}. This
complicates the algorithmic approach, since both types of paths must be
controlled simultaneously. Consider Fig.~\ref{fig:RNA-two-graphs}.  The
shortest path between the green and blue regions includes some vertices
outside the interval between these two regions. The basic idea is to
generalize Equation~(\ref{eq:ZI}) to computing the partition function
$Z^{v,w}[d]$. The main question now becomes how to recurse over
decompositions of both the inside and the outside paths.

Fig.~\ref{fig:RNA-two-graphs} shows that the outside paths are important
for the green region, i.e., the region that is covered by an arc. Hence, we
have to consider the different cases that the two positions $v$ and $w$ are
covered by arcs. The set $\Omega$ of all secondary structures on $\xi$ can
be divided into two disjoint subclasses that have to be treated
differently:
\begin{itemize}
  \item[$\Omega_0$] $v$ and $w$ are not enclosed in a common base pair,
    i.e., $B_v\cap B_w=\emptyset$.
  \item[$\Omega_1$] there is a base pair enclosing both $v$ and $w$, i.e.,
   $B_v\cap B_w\ne\emptyset$. 
\end{itemize}
Note that this bipartition explicitly depends on $v$ and $w$.  In the
following, we will first introduce the recursions that are required in
$\Omega_0$ structures to compute $Z^{v,w}[d]$.

\subsubsection{Contribution of $\Omega_0$ structures to $Z^{v,w}[d]$}
\label{S:omega0} 
One example of this case is given in Fig.~\ref{fig:RNA-two-graphs} with
the red and blue region, where $v$ (vertex in green region) is covered by
an arc, and $w$ (vertex in blue region) is external.  Denote the
$\prec$-maximal base pair enclosing $v$ by $\{i,j\}$. Since at most one of
$v$ and $w$ is covered by an arc, we know that $j<w$. Hence, every path $p$
from $v$ to $w$, and hence also the shortest paths (not necessarily unique)
must run through the right end $j$ of the arc $\{i,j\}$. More precisely,
there must sub-paths $p_1$ and $p_2$ with $d(p)=d(p_1)+d(p_2)+a$ such that
$
  v \stackrel{p}{\spth} w  \to   v \stackrel{p_1}{\spth} j - (j+1) 
  \stackrel{p_2}{\spth} w 
$,
where $i \stackrel{p}{\spth} j$ denotes that $p$ is \textbf{a} shortest 
path from $i$ to $j$ and $-$ denotes a single backbone edge. For the
shortest path from $v$ to $j$, it consists either of 
a shortest path $v \stackrel{p'}{\spth} i$ and the arc $\{i,j\}$,
or it goes directly to $j$ without using the arc $\{i,j\}$.

How does this distinction translate to the partition function approach? 
If we want to calculate the contribution of this case to the
partition function $Z^{v,w}[d]$, we have to split both the
sequence $\xi[i,w]$ and distance $d$ as follows 
\begin{center}
a.)\quad{} \includegraphicstop[width=0.5\textwidth]{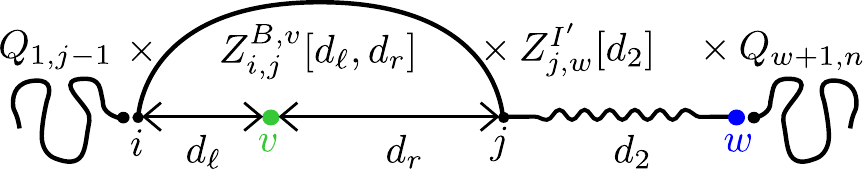}
\end{center}
where $\ZIprime_{j,w}[d_2]$ is the partition function
starting and ending with a single-stranded base as defined in
Equation~(\ref{eq:ZIprime}), and 
$\ZB{v}_{i,j}[d_\ell,d_r]$ is the partition function consisting of all
structures of $\xi[i,j]$ containing the base pair $\{i,j\}$ with the
property that the shortest path from $v$ to $i$ has length $d_\ell$ and the
shortest path from $v$ to $j$ has length $d_r$. In addition,
$d$, $d_r$ and $d_2$ must satisfy $d=d_r+d_2$. 

The remaining cases for the contribution of the class $\Omega_0$ to
$Z^{v,w}[d]$ are given by all other possible combinations of $v$ and $w$ being
single-stranded or being covered by an arc, i.e.,
\begin{center}
  \includegraphics[width=1\textwidth]{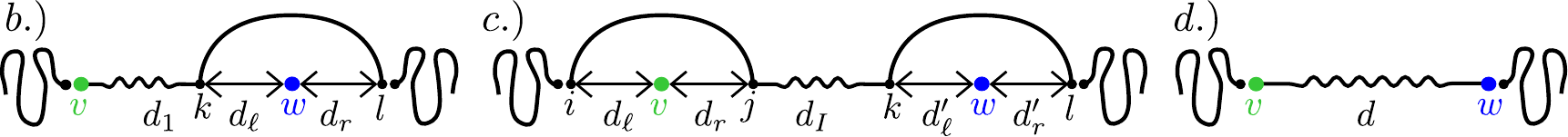}
\end{center}
To simplify, we extend the definition of $\ZB{v}_{i,j}[d_\ell,d_r]$
by setting $\ZB{v}_{v,v}[0,0]=1$ and $\ZB{v}_{v,v}[d_\ell,d_r]=0$ for
$d_\ell +d_r>0$. This allows us to conveniently model all cases where
either $v$ or $w$ are external, i.e., a.), b.), and d.), as special 
cases of c.).

In case c.) we have to split the distance $d$ into four contributions and
we require two splitting positions for the sequence for all combinations of
$i,j,v,w$. This would result in an $O(n^{6}D^5)$ algorithm. A careful
inspection shows, however, that the split of the distances for the arcs
into $d_\ell$ and $d_r$ is unnecessary. Since we want to know only distance
to the left/right end overall, we can simply introduce two matrices
$\ZB{v,\ell}_{i,j}[d]$ and $\ZB{v,r}_{i,j}[d]$ that store these
values. These matrices can be generated from $\ZB{v}_{i,j}[d_\ell,d_r]$ as
follows:
\begin{displaymath}
\ZB{v,\ell}_{i,j}[d] =\sum_{\substack{d_r\\ d_r+b\geq d}}
\ZB{v}_{i,j}[d,d_r] + 
\sum_{\substack{d_\ell\\ d_\ell > d}} \ZB{v}_{i,j}[d_\ell,d-b]
\end{displaymath}
Analogously, we compute $\ZB{v,r}_{i,j}[d]$.

Overall, the contribution to $Z^{v,w}[d]$ for structures in $\Omega^0$ is
given by
\begin{equation}
\label{eq:Zomega0}
Z_0^{v,w}[d] =   
\sum_{\substack{d_1,d_2\\
d_1+d_2\leq d
}}\sum_{\substack{i,j,k,l\\ 
i \leq v\leq
  j<k\leq w\leq l}}
\left(
\begin{array}{l}
Q_{1,i-1}\cdot \ZB{v,r}_{i,j}[d_1] \\
\cdot\,\ZIprime_{j,k}[d-(d_1+d_2)]\\
\cdot\, \ZB{w,\ell}_{k,l}[d_2]\cdot Q_{l+1,n}
\end{array}
\right)
\end{equation}
Note that for splitting the distance, we reuse the same indices (e.g., the
$\Red{j}$ in
$\ZB{v,r}_{i,¸\Red{j}}[d_1]\cdot\ZIprime_{\Red{j},k}[d-(d_1+d_2)]$, where
as for the remaining partition function, we use successive indices
(e.g.,the $\Red{i}$ in $Q_{1,\Red{i-1}}\cdot
\ZB{v,r}_{\Red{i},j}[d_1]$). This difference comes from the fact that
splitting a sequence into subsequences is done naturally between two
successive indices, whereas splitting a distance is naturally done by
splitting at an individual position. We have only to guarantee that the
substructures which participate in the split do agree on the structural
context of the split position. This is guaranteed by requiring that
$\ZIprime$ starts and ends with a backbone edge. We note that the
incorporation of the full dangling end parameters makes is more tedious to
handle the splitting positions.

This results in a complexity of $O(n^6D^3)$ time and $O(n^3D)$
space. However, we do not need to split in $i,j,k,j$
simultaneously. Instead, we could split case (c) at position $j$ and 
introduce for all $v\leq j$ and $k\leq w$ the auxiliary variables
\begin{eqnarray*}
Z_{1,j}^{B,v,r}[d_1] &=& \sum_{i \leq v} 
Q_{1,i-1}\cdot
  \ZB{v,r}_{i,j}[d_1]\ \ \ \ \ 
\ZB{w,\ell}_{k,n}[d_2]= \sum_{w\leq l}\ZB{w,\ell}_{k,l}[d_2]\cdot
Q_{l+1,n}\\
 Z^{I\!B,w,\ell}_{j,n}[d']&=& \sum_{k >j}\sum_{\substack{d_2\leq d'
  }}\ZIprime_{j,k}[d'-d_2]\cdot\ZB{w,\ell}_{k,n}[d_2].
\end{eqnarray*}
Finally, we can replace recursion~$(\ref{eq:Zomega0})$ by
\begin{eqnarray}
\label{eq:Zomega0-eff}  Z_0^{v,w}[d] &=& \sum_{v\leq j} \sum_{d_1 \leq d} Z_{1,j}^{B,v,r}[d_1] \cdot
  Z^{I\!B,w,\ell}_{j,n}[d-d_1]
\end{eqnarray}
We thus arrive at $O(n^4D^2)$ time and $O(n^3D)$ space complexity for the
contribution of $\Omega_0$ structures to $Z^{v,w}[d]$, excluding the
complexity of computing $\ZB{v}_{i,j}[d_\ell,d_r]$.

\subsubsection{Contribution of $\Omega_1$ structures to $Z^{v,w}[d]$}
\label{S:omega1}

$\Omega_1$ contains all cases where $v$ and $w$ are covered by a base
pair. In the following, let $\{p,q\}$ be the $\prec$-minimal base pair
covering $v$ and $w$.  In principle, this case looks similar to the overall
case for $\Omega_0$. However, we have now to deal not only with an inside
distance, but also with an outside distance over the base pair
$\{p,q\}$. Thus, we need to store the partition function for all inside and
outside for each $\prec$-minimal arc $\{p,q\}$ that covers $v$ and $w$,
which we will call $\YB{v,w}_{p,q}[d_O,d_I]$.  In principle, a similar
recursion as defined for $Z_0$ in equation~$(\ref{eq:Zomega0})$ can be
derived, with the additional complication since we have to take care of the
additional outside distance due to the arc $(p,q)$. Thus, we obtain the
following splitting:
\begin{center}
  \includegraphics[width=0.7\textwidth]{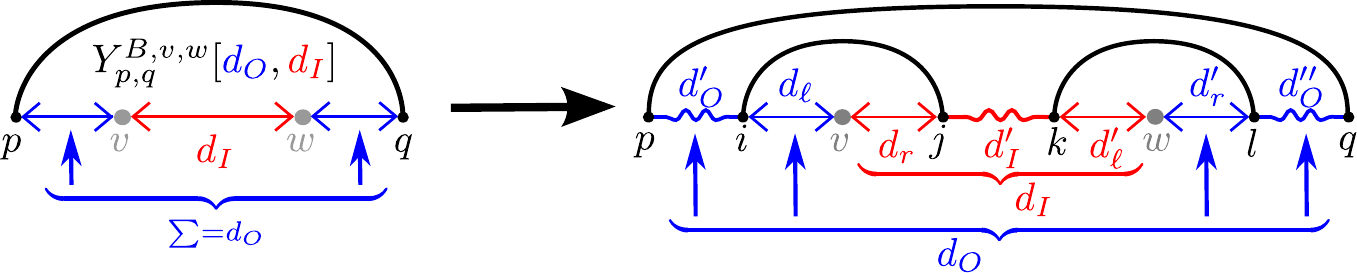}
\end{center}
Again we can avoid the complexity of simultaneously  splitting at
$\{i,j\}$ \emph{and} $\{k,l\}$ by doing a major split after $j$. Thus, we
get the equivalent recursions as in 
eqns.(\ref{eq:omega1-case1}--\ref{eq:omega1-case3}):
\begin{eqnarray}
\label{eq:omega1-case1}  Y_{p,j}^{B,v,r}[d,d_r] &=& \sum_{p<i \leq v} 
\sum_{d_O'\leq d}\ZIprime_{p,i}[d_O']\cdot
  \ZB{v}_{i,j}[\overbrace{d-d_O'}^{\hat{=}\,d_\ell},d_r]\\
\label{eq:omega1-case2}
\YB{w,\ell}_{k,q}[d_\ell',d]&=& \sum_{w\leq l <q}\sum_{d_O''\leq
  d}\ZB{w}_{k,l}[d_\ell',\overbrace{d-d_O'}^{\hat{=}\,d_r'}]\cdot \ZIprime_{l,q}[d_O'']\\
\label{eq:omega1-case3} 
Y^{I\!B,w,\ell}_{j,q}[d_I',d]&=& \sum_{j <k < q}\sum_{\substack{d_\ell'\leq d_I'
  }}\ZIprime_{j,k}[d_I'-d_\ell']\cdot\YB{w,\ell}_{k,q}[d_\ell',d]
\end{eqnarray}
Overall, we get the following recursion:
\begin{eqnarray}
\label{eq:Zomega1-eff}  Z_{p,q}^{v,w}[d_O,d_I] &=& \sum_{v\leq j}
\sum_{\substack{d_r \leq d_I\\
  d \leq d_O}} Y_{p,j}^{B,v,r}[d,d_r] \cdot 
Y^{I\!B,w,\ell}_{q,j}[d_I-d_r,d_O-d]
\end{eqnarray}

Overall, we can now define $Z^{v,w}[d]$ by
\begin{eqnarray*}
  Z^{v,w}[d] &=& Z_0^{v,w}[d] 
  + \sum_{\substack{\{p,q\} \neq \{v,w\}\\[1.0ex] {d_I\geq d+b}}}
  Z_{p,q}^{v,w}[d,d_I]
  + \sum_{\substack{\{p,q\} \neq \{v,w\}\\[1ex]{d < d_O+b}}} 
    Z_{p,q}^{v,w}[d_O,d]
\end{eqnarray*}
This part has now a complexity of $O(n^4D^2)$ space and $O(n^5D^4)$
time. For practical applications, however, we do not need to consider all
possible $\{p,q\}$. Instead, there are only few base pairs that are likely
to form \emph{and} that cover $v,w$, especially for $v,w$ where the
internal distance of $v,w$ is large enough such that an outside path has to
be considered at all. If we assume a constant number of such long-range
base-pairs, then the complexity is reduced by an $n^2$-factor. For the
complexity in terms of distance, recall that $D$ is typically small.

\subsection{Recursions for $\ZB{v}_{i,j}[d_\ell,d_r]$}
\label{S:Part3}

So far, we have used $\ZB{v}_{i,j}[d_\ell,d_r]$ as a black box. In order to 
compute these terms, we distinguish the limiting cases a.) $v=i$, 
b.) $v=j$, c.) is external from the generic case d.): 
\begin{center}
  \includegraphics[width=\textwidth]{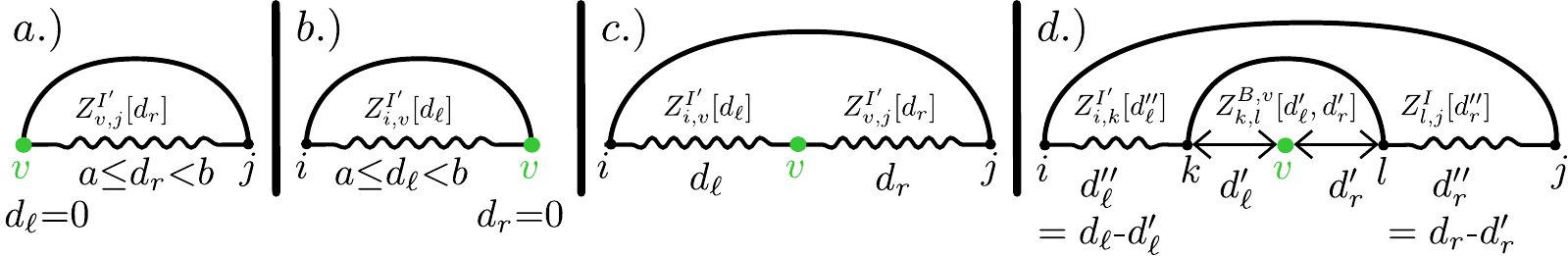}
\end{center}
Starting from the limiting cases, we initialize 
$\ZB{v}_{v,j}[0,d_r]$ as follows:
\begin{eqnarray*}
\ZB{v}_{v,j}[0,d_r]=
\begin{cases}
 \ZIprime_{v,j}[d_r]  &\text{for $a\leq d_r < b$}\\
 \sum_{d'\geq b}\ZIprime_{v,j}[d'] & \text{for $d_r = b$}\\
 0 & \text{otherwise}
\end{cases}
\end{eqnarray*}
and analogously for $\ZB{v}_{i,v}[d_\ell,0]$. Furthermore,
 $\ZB{v}_{i,j}[0,0] = 0$ for $i\neq v\neq j$. Finally, we have the
 following recursion for $i\neq v \neq j$, $d_\ell>0$ and $d_r>0$:
\begin{eqnarray}
\label{eq:ZB-overall-full}  \ZB{v}_{i,j}[d_\ell,d_r] &=&
\widehat{Q}^b_{i,j}\cdot
\sum_{\substack{k\neq l\\ i<k\leq v\\ v\leq l<j}}
\sum_{\substack{d_\ell'\leq d_\ell\\ d_r'\leq d_r}} 
\ZIprime_{i,k}[d_\ell-d_\ell']\cdot 
\ZB{v}_{k,l}[d_\ell',d_r']\cdot \ZIprime_{l,j}[d_r-d_r']
\end{eqnarray}
where $\widehat{Q}^b_{i,j}$ is the external partition function over all
structures on the union of the intervals $\xi[1..i]\cup\xi[j..n]$ so that
$\{i,j\}$ is a base pair. This is equivalent to
$\widehat{Q}^b_{i,j}=Pr(\{i,j\})\times Q/Q^{b}_{i,j}$.
The base pair probability $Pr(\{i,j\})$, and the partition functions 
$Q$ and $Q^{b}_{i,j}$ are computed by means of McCaskill's algorithm.

Recursion (\ref{eq:ZB-overall-full}) apparently has complexity $O(n^5D^4)$
in time and $O(n^3D^2)$ in space.  This can be reduced due to the strong
dependency between $d_\ell$ and $d_r$, however.  By construction we have
$|d_\ell-d_r| \leq b$ since we can always use the bond $\{i,j\}$ to
traverse from one end to the other. Furthermore, assuming integer values
for $a$ and $b$, we can have only $c_b=2 b/\lcd(a,b)+1$ different values
for $(d_\ell-d_r)$ This implies that the space complexity of
$\ZB{v}_{i,j}[d_\ell,d_r]$ is $O(n^3 D c_b)$. Instead of 
$\ZB{v}_{i,j}[d_\ell,d_r]$, we store
$\ZB{v}_{i,j}[d_\ell,d_\ell+d_{\text{add}}]$ for the $c_b$ possible values
of $d_{\text{add}}$.

The dependency between $d_\ell$ and $d_r$ can also be used to reduce the
time complexity in Equ.(\ref{eq:ZB-overall-full}). The problematic case
is (d).  Instead of using the variables $d_\ell$ and $d_r$ in
$\ZB{v}_{i,j}[d_\ell,d_r]$ we use the pair $d_\ell,d_{\text{add}}$ in
$\ZB{v}_{i,j}[d_\ell,d_\ell+d_{\text{add}}]$. Similarly, we use
$d_\ell',d_{\text{add}}'$ instead of $d_\ell',d_r'$ for the inner base
pair, which then determines completely the splitting the distances. The
details are relegated to Appendix C. Overall, this results in an recursion
for $\ZB{v}_{i,j}[d_\ell,d_\ell+d_{\text{add}}]$ with complexity
$O(n^5c_b^2)$ time and $O(n^3Dc_b)$ space.

\section{Discussion and Applications}
The theoretical analysis of the distance distribution problem shows that,
while polynomial-time algorithms exist, they probably cannot the improved
to space and time complexities that make them widely applicable to large
RNA molecules. Due to the unfavorable time complexity of the current
algorithm and the associated exact implementation in C, a rather simple and
efficient sampling algorithm has been implemented.  We resort to sampling
Boltzmann-weighted secondary structures with \texttt{RNAsubopt -p}
\cite{Lorenz:11a}, which uses the same stochastic backtracing approach as
\texttt{sfold} \cite{Ding:03}. As the graph-distance for a pair of
nucleotides in a given secondary structure can be computed in $O(n\log n)$
time, even large samples can be evaluated efficiently\footnote{The C++
  program \texttt{RNAgraphdist} is available from
  \url{http://www.rna.uni-jena.de/supplements/RNAgraphdist/RNAgraphdist1.0.tar.gz}.}.

\begin{figure}[t]
  \centering
    \includegraphics[width=\textwidth]{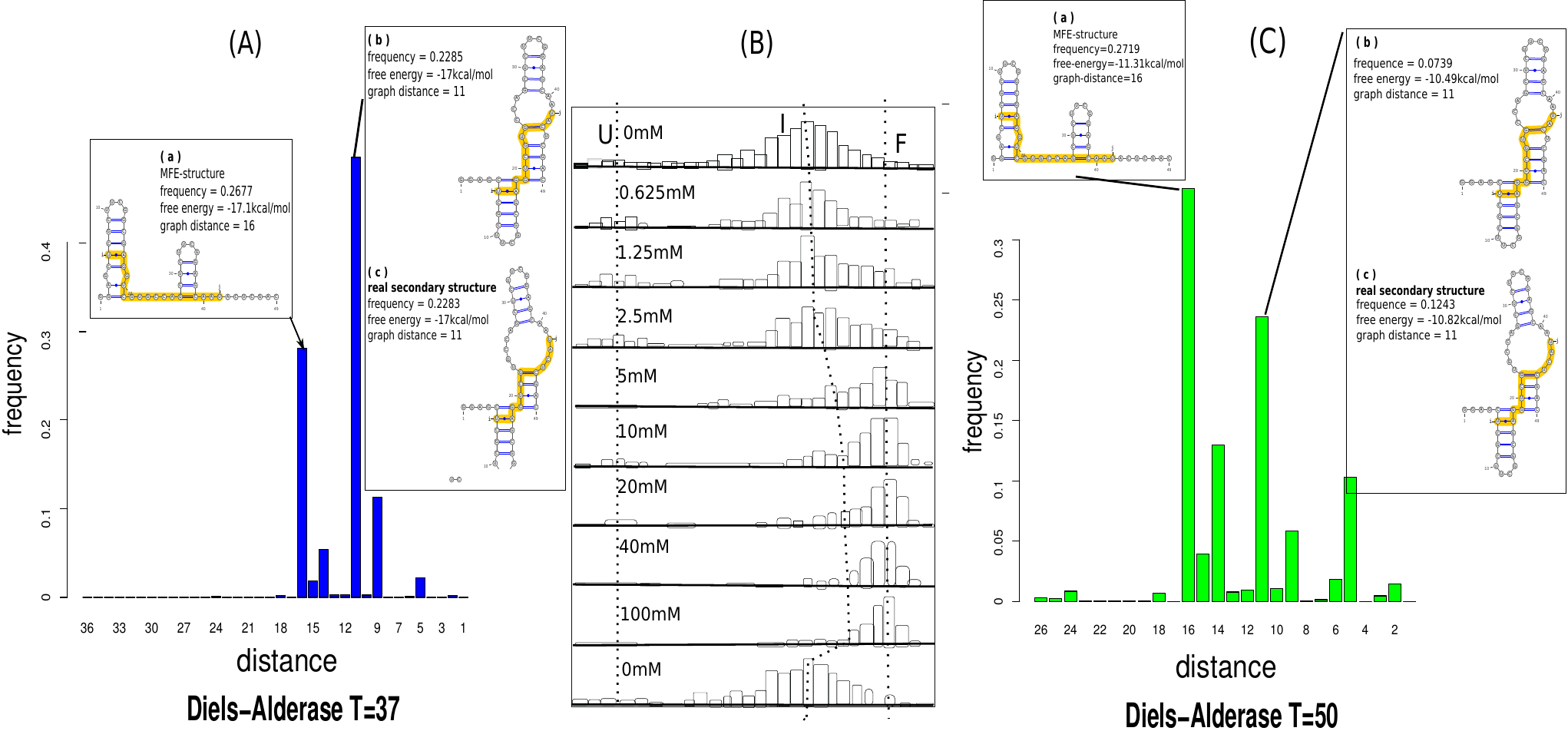}
    \caption{Relation between graph distance distribution and smFRET
      data. (A) The graph distance distribution of a Diels-Alderase
      ribozyme at
      temperature 37$^\circ$C.. Structures (a), (b) and (c) are the top three secondary
      structures considering their free energy. In which, the minimum free
      energy structure is showed in (a), (c) is the real secondary
      structure which is ranked as the 3rd best sub-optimal structure with
      \texttt{RNAsubopt -e}. he graphic representations of these structures
      are produced with \texttt{VARNA} \cite{varna}.  (B) The corresponding
      smFRET efficiency ($E_{fret}$) histograms are reported in
      \cite{Kobitski:07}.  From these data, three separate states of the
      DAse ribozyme can be distinguished, the unfolded (U), intermediate
      (I) and folded (F) states. (C) The graph distance distribution in the
      ensemble which is approximated with \texttt{RNAsubopt -p} at
      temperature 50$^\circ$C.}
  \label{fig:dase}
\end{figure}

As we pointed out in the introduction, the graph distance measure
introduced in this paper can serve as a first step towards a structural
interpretation of smFRET data. As an example, we consider the graph
distance distribution of a Diels-Alderase (DAse) ribozyme
(Fig.~\ref{fig:dase} (A)). Histograms of smFRET efficiency ($E_{fret}$) for
this 49 nt long catalytic RNA are reported in \cite{Kobitski:07} for a
large number of surface-immobilized ribozyme molecules as a function of the
Mg$^{2+}$ concentration in the buffer solution. A sketch of their
histograms is displayed in Fig.~\ref{fig:dase} (B).  The dyes are attached
to sequence positions 6 (Cy3) and 42 (Cy5) and hence do not simply reflect
the end-to-end distance, Fig.~\ref{fig:dase} (A)(c).  In this example, we
observe the the expected correspondence small graph distances with a strong
smFRET signal.  This is a particular interesting example, since the minimal
free energy (mfe) structure (Fig.~\ref{fig:dase} (A)(a)) predicted with
\texttt{RNAfold} is not identified with the real secondary structure
(Fig.~\ref{fig:dase} (A)(c)). In fact, the ground state secondary structure
is ranked as the 3rd best sub-optimal structure derived via
\texttt{RNAsubopt -e}. The free energy difference between these two
structures is only $0.1kcal/mol$. However, their graph distances show a
relatively larger difference. The 2nd best sub-optimal structure
(Fig.~\ref{fig:dase} (A)(b)) looks rather similar with the 3rd structure,
in particular, they share the same graph distance value.

The smFRET data of \cite{Kobitski:07} indicate the presence of three
sub-populations, corresponding to three different structural states: folded
molecules (state F), intermediate conformation (state I) and unfolded
molecules (state U).  In the absence of Mg$^{2+}$, the I state dominates,
and only small fractions are found in states U and F. Unfortunately, the
salt dependence of RNA folding is complex \cite{Leipply:09,Mathews:99} and
currently is not properly modeled in the available folding programs. We
can, however, make use of the qualitative correspondence of low salt
concentrations with high temperature. In Fig.~\ref{fig:dase} (C) we
therefore re-compute the graph distance distribution in the ensemble at an
elevated temperature of 50$^\circ$C. Here, the real structure becomes the
second best structure with free energy $-10.82kcal/mol$ and we observe a
much larger fraction of (nearly) unfolded structures with longer distances
between the two beacon positions. Qualitatively, this matches the smFRET
data showed in Fig.~\ref{fig:dase} (B).
 
\begin{figure}[t]
  \centering
    \includegraphics[width=0.48\textwidth]{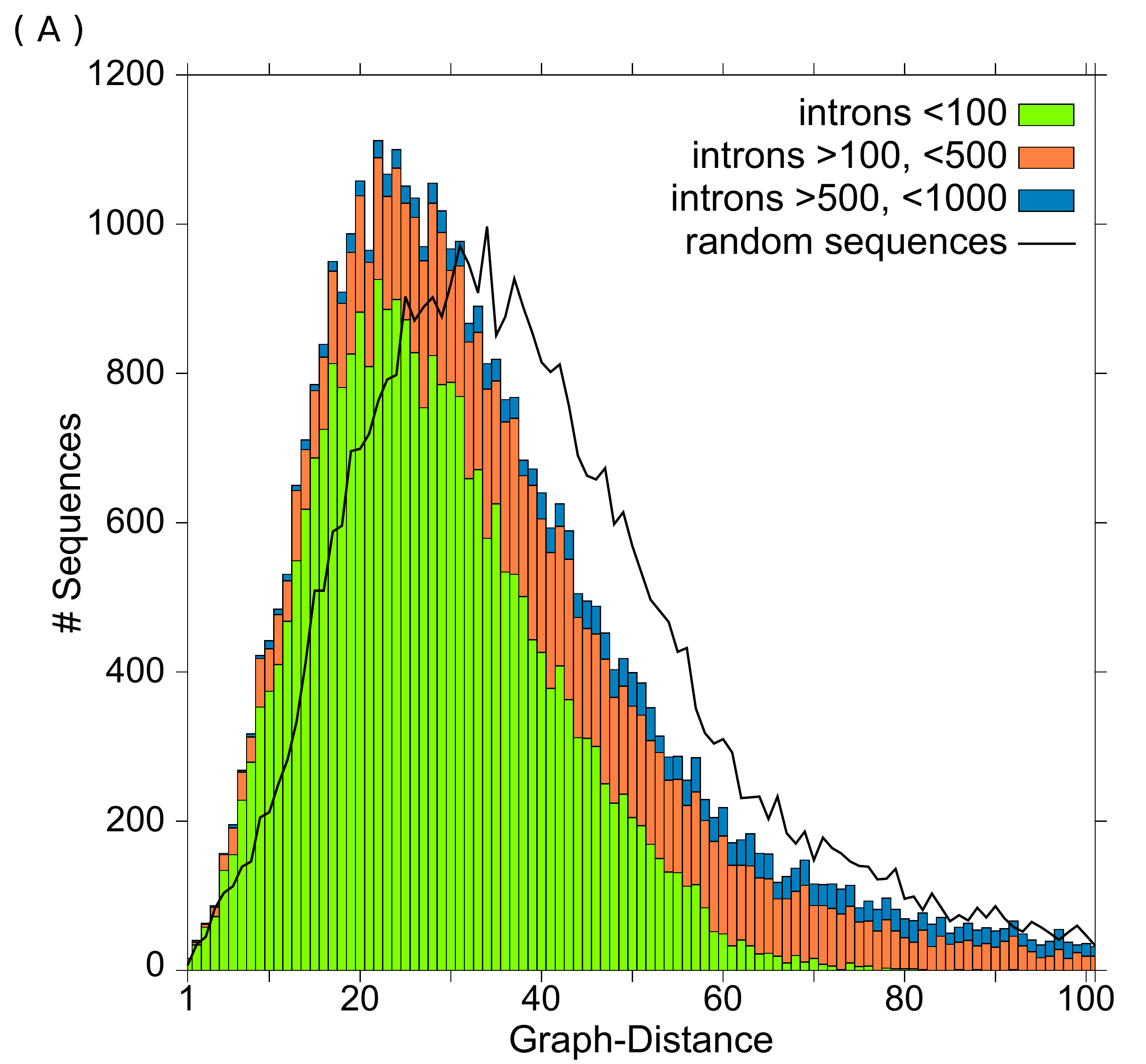} 
    \includegraphics[width=0.48\textwidth]{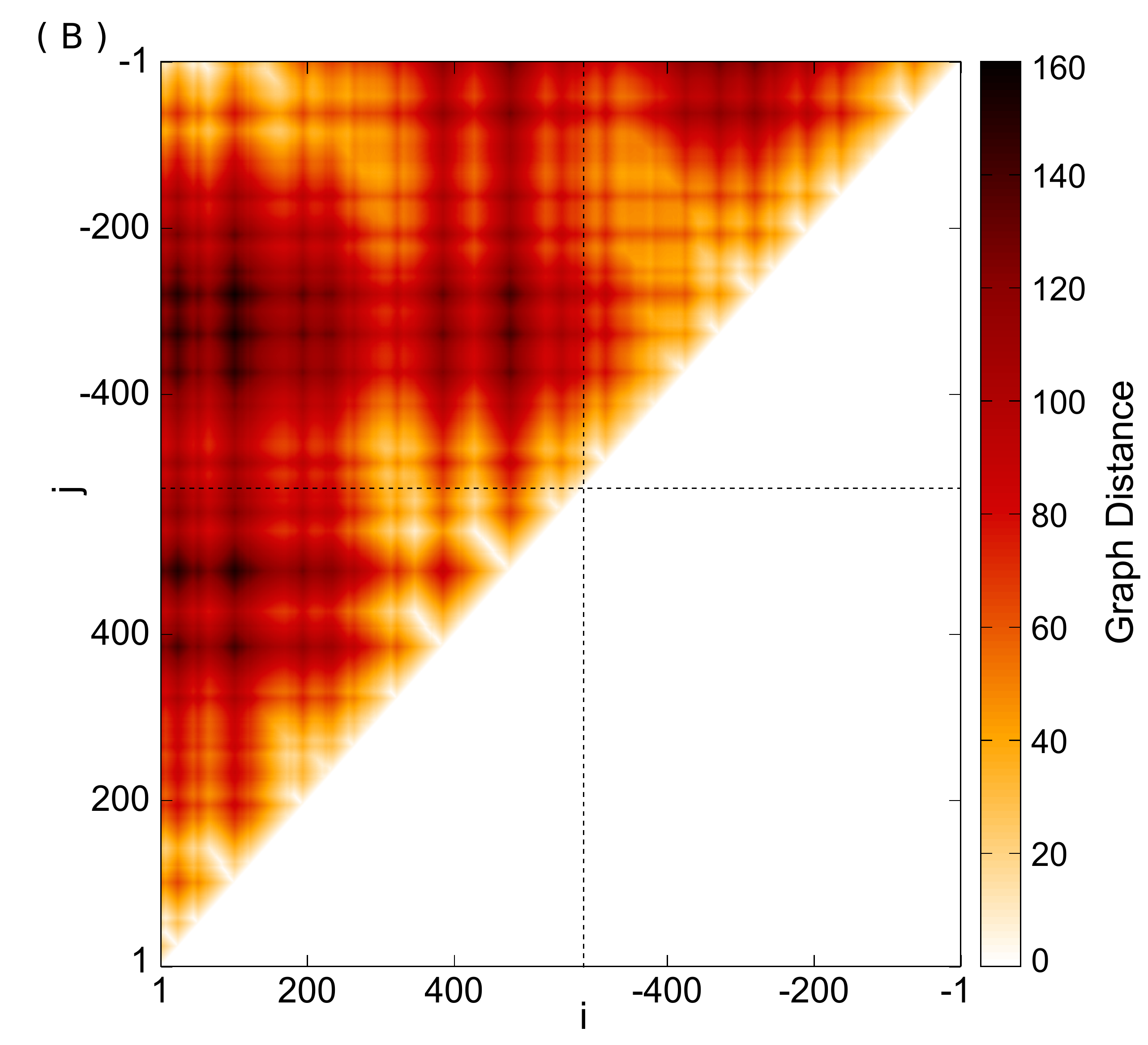} 
    \caption{{\bf(A)}: Distribution of graph-distances ($a=b=1$) in
      \emph{Drosophila melanogaster} pre-mRNAs between the first and last
      intron position.  To save computational resources, pre-mRNAs were
      truncated to 100 nt flanking sequence. 
      The black curve shows the graph-distance distribution computed for the 
      corresponding pairs of positions on sequences that were randomized by 
      di-nucleotide shuffling.
      {\bf (B)}: Graph-distances ($a=b=1$) within and between the
      5' and 3' regions of the genomic RNA of human \emph{Coronavirus} 229E
      computed from a concatenation of position 1--576 and
      25188--25688. Secondary structures bring the 5' TRS-L and 3' TRS-B
      elements into close proximity. More detailed information related
      to this example can be found in Supplemental Material D.}
  \label{fig:examples}
\end{figure}

Long-range interactions play an important role in pre-mRNA splicing and in
the regulation of alternative splicing \cite{Baraniak:2003,McManus:11},
bringing splice donor, acceptor, branching site into close spatial
proximity. Fig.~\ref{fig:examples}(A) shows for \emph{D.\ melanogaster}
pre-mRNAs that the distribution of graph-distances between donor and
acceptor sites shifted towards smaller values compared to randomly selected
pairs of positions with the same distance. \footnote{Due to the
  insufficiency of the spacial-distance information of structural elements
  in the secondary structures, we artificially choose $a=b=1$ in our
  experiments.}  Although the effect is small, it shows a clear difference
between the real RNA sequences and artificial sequences that were
randomized by di-nucleotide shuffling.

The spatial organization of the genomic and sub-genomic RNAs is important
for the processing and functioning of many RNA viruses. This goes far
beyond the well-known panhandle structures. In \emph{Coronavirus} the
interactions of the 5' TRS-L cis-acting element with body TRS elements has
been proposed as an important determinant for the correct assembly of the
\emph{Coronavirus} genes in the host \cite{Dufour:11}. The matrix of
expected graph-distances in Fig.~\ref{fig:examples}(B) shows that TRS-L and
TRS-B are indeed placed near each other. More detailed information is
provided in Appendix (D).

Our first results show that the systematic analysis of the graph-distance
distribution both for individual RNAs and their aggregation over ensembles
of structures can provide useful insights into structural influences on RNA
function. These may not be obvious directly from the structures due to the
inherent difficulties of predicting long-range base pairs with sufficient
accuracy and the many issues inherent in comparing RNA structures of very
disparate lengths.

Due the complexity of algorithm we have refrained from attempting a direct
implementation in an imperative programming language. Instead, we are
aiming at an implementation in Haskell that allows us to make use of the
framework of algebraic dynamic programming \cite{Giegerich:02}. 
The graph distance measure and the associated algorithm can
be extended in principle to of RNA secondary structures with additional
tertiary structural elements such as pseudoknots \cite{Nebel:11} and
G-quadruples \cite{Lorenz:13}.  RNA-RNA interaction structures
\cite{ripalign} also form a promising area for future extensions. We note
finally, that the Fourier transition method introduced in
\cite{senter:2012} could be employed to achieve a further speedup.

\par\noindent\textbf{Acknowledgements.} This work was supported in part by 
the \emph{Deutsche Forschungsgemeinschaft} proj.\ nos.\ BA 2168/2-2, STA
850/10-2, SPP 1596 and MA5082/1-1.

\bibliographystyle{plain}
\bibliography{distance}


\end{document}